# Accelerated Discovery of Crystalline Materials with Record Ultralow Lattice Thermal Conductivity via a Universal Descriptor

Xingchen Shen[1,2,#,*], Jiongzhi Zheng[3,#], Michael Marek Koza[4], Petr Levinsky[5], Jiri Hejtmanek[5], Philippe Boullay[2], Bernard Raveau[2], Jinghui Wang[6], Jun Li,[6] Pierric Lemoine[7], Christophe Candolfi[7], Emmanuel Guilmeau[2,*]

[1]*MOE Key Laboratory of Materials Physics and Chemistry under Extraordinary Conditions, School of Physical Science and Technology, Northwestern Polytechnical University, Xi'an, 710072 China*

[2]*CRISMAT, CNRS, Normandie Univ, ENSICAEN, UNICAEN, 14000 Caen, France*

[3]*Thayer School of Engineering, Dartmouth College, Hanover, New Hampshire 03755, USA*

[4]*Institute Laue Langevin, 71 avenue des Martyrs CS 20156, Grenoble, Cedex 9, 38042 France*

[5]*FZU - Institute of Physics of the Czech Academy of Sciences, Cukrovarnická 10/112, Prague 6 16200, Czech Republic*

[6]*ShanghaiTech Laboratory for Topological Physics & School of Physical Science and Technology, ShanghaiTech University, Shanghai 201210, China*

[7]*Université de Lorraine, CNRS, IJL, F-54000 Nancy, France*

[#] *These authors contributed equally to this work*

[*]*Corresponding authors*: [xingchen.shen@nwpu.edu.cn,](xingchen.shen@nwpu.edu.cn)

[emmanuel.guilmeau@ensicaen.fr](emmanuel.guilmeau@ensicaen.fr)

## ABSTRACT

Ultralow glass-like lattice thermal conductivity in crystalline materials is crucial for enhancing energy conversion efficiency in thermoelectrics and thermal insulators. We introduce a universal descriptor for thermal conductivity that relies only on the atomic number in the primitive cell and the sound velocity, enabling fast and scalable materials screening. Coupled with high-throughput workflows and universal machine learning potentials, we identify the candidate materials with ultralow thermal conductivity from over 25, 000 materials. We further validate this approach by experimentally confirming record-low thermal conductivity values of 0.15–0.16 W/m·K from 170 to 400 K in the halide metal $CsAg_2I_3$. Combining inelastic neutron scattering with first-principles calculations, we attribute the ultralow thermal conductivity to the intrinsically small sound velocity, strong anharmonicity, and structural complexity. Our work illustrates how a universal descriptor, combined with high-throughput screening, machine-learning potential and experiment, enables the efficient discovery of materials with ultralow thermal conductivity.

## Introduction

The discovery of crystalline materials with ultralow glass-like lattice thermal conductivity (ULG-type $\kappa_L$) is essential for maximizing performance in thermoelectrics,[1] thermal insulators,[2] and thermal coating barriers.[3] Generally, crystalline materials with complex structures, large primitive cells (containing atoms > 20), and heavy elements tend to exhibit low $\kappa_L$.[4] However, ultralow $\kappa_L$[5] can also occur in materials with small primitive cells and light elements, which is due to the presence of rattler atoms,[6] strong lattice anharmonicity,[7, 8] or bonding heterogeneity.[9] Recently, significant progress has been made in both experimental and theoretical investigations aimed at uncovering crystalline materials with very low $\kappa_L$ below 1.0 W/m·K. Experimentally, Mukhopadhyay et al.[10] developed a class of thallium selenides and measured an ultralow $\kappa_L$ of approximately 0.30 W/m·K at 300 K in $Tl_3VSe_4$. This value was attributed to its exceptionally low group velocity and strong anharmonicity. Acharyya et al.[11] synthesized layered halide perovskite $Cs_3Bi_2I_6Cl_3$ single crystal and measured an ultralow thermal conductivity of 0.20 W/m·K along in-/cross-plane directions at 300 K, which was attributed to its low average sound velocity ($v$) and weak chemical bonds. Gibson et al.[12] found that the 2D layered structure of mixed anion $Bi_4O_4SeCl_2$ with specific spatial arrangement of interfaces suppresses acoustic phonon contributions, achieving an ultralow $\kappa_L$ of 0.10 W/m·K in the cross-plane direction at 300 K. Interestingly, for the even more complex crystal structure of argyrodite $Cu_7PS_6$ with 54 atoms in primitive cell, Shen et al.[13] reported an ultralow $\kappa_L$ of 0.50 W/m·K at 300 K. Notably, these crystalline materials with ultralow $\kappa_L$ also generally exhibit a weak/positive temperature-dependent behavior, resembling ULG-type thermal conduction. These experimental observations of ULG-type $\kappa_L$ raise an open question: how can $\kappa_L$ be pushed to its lower limit over a broad temperature range based on material properties, such as anharmonicity and structural complexity?

Theoretically, to accurately explain the experimental ULG-type $\kappa_L$, Simoncelli *et al.*[14] proposed a unified theory of thermal transport that considers both particle-like phonon propagation (phonon contribution) and wave-like tunneling channels (diffuson contribution), successfully reproducing the glass-like temperature dependence of $\kappa_L$ in crystalline $CsPbBr_3$. Building on the unified theory of thermal transport, Xia *et al.*[15] further incorporated the effects of anharmonic temperature dependence and higher-order phonon scattering, successfully replicating the temperature-independent ultralow $\kappa_L$ in crystalline $Cu_{12}Sb_4S_{13}$. Moreover, the well-established unified theory of thermal transport has been successfully applied to explain thermal transport in various materials with complex structures or strong anharmonicity, such as $Yb_{11}M_4Sb_{11}$ (M=Mg and Mn),[16] $Cu_7PS_6$,[13] $NaAg_3S_2$,[17] $K_2Ag_4Se_3$,[18] $YbFe_4Sb_{12}$,[19] and $MAPbI_3$ (MA=methylammonium).[20] Therefore, exploring materials with ULG-type $\kappa_L$, approaching the glassy limit in crystalline compounds, is not only fundamentally intriguing but also essential for understanding the complex interplay between crystal structure, bonding strength, and anharmonic lattice dynamics. These investigations offer valuable criteria for identifying unknown low-$\kappa_L$ crystalline materials and open new avenues for engineering heat transport properties in existing compounds.

Despite significant advancements in experimental observations and theoretical predictions of $\kappa_L$ in crystalline materials, the high cost remains a major obstacle for high-throughput calculations/experiments and rapid screening of crystalline materials with ULG-type $\kappa_L$. To simplify the theoretical prediction process for identifying materials with ultralow $\kappa_L$, Knoop *et al.*[21] introduced the concept of the degree of anharmonicity (DOA), derived from *ab initio* molecular dynamics (AIMD), to quantitatively evaluate $\kappa_L$ in crystalline materials. Using the concept of DOA, Zeng *et al.*[22] theoretically and experimentally reported an ultralow $\kappa_L$ of 0.25 W/m·K at 300 K in the simple crystalline material $AgTlI_2$, attributed to its Ag-associated strong anharmonicity. Additionally, considering the inverse relationship between population/diffuson thermal conductivity and number of atoms in a primitive cell ($n$), they proposed a conceptual diagram to facilitate the search for ultralow $\kappa_L$ crystalline materials with simple crystal structure. Although the AIMD-based DOA concept facilitates the rapid screening of materials with ultralow $\kappa_L$, the lengthy AIMD calculations remain computationally expensive, and the DOA concept is limited to qualitatively evaluating thermal transport properties.

Over the past decades, widely known empirical thermal transport models, such as the Slack[23] and Debye-Callaway[24] models, have been proposed to estimate the thermal conductivity of crystalline materials. However, these models are limited in accuracy and rely heavily on numerous empirical parameters. To overcome the limitations of the aforementioned empirical models, several proposed an improved and straightforward minimum thermal conductivity model based on diffuson-mediated thermal transport theory. Furthermore, inspired by the unified theory of thermal transport, Xia *et al.*[25] recently proposed a minimum thermal conductivity model that incorporates contributions from both propagons and diffusons. Meanwhile, Wang *et al.*[26] derived an interpretable formula for $\kappa_L$ of crystals by utilizing a large database and building upon the Slack model. Although recently developed thermal transport models have achieved significant success, their complexity[25, 26] or limited prediction accuracy[27] hinders their practical application by experimentalists. A simple unified thermal transport indicator based on experimentally accessible parameters has still not yet been developed, posing a significant challenge to the efficient exploration of crystalline materials with desirable thermal transport properties. In addition, with the rapid advancements in universal machine learning force fields (MLFF), such as MACE,[28] M3GNET,[29] CHGNET,[30] and MatterSim,[31, 32] integrating MLFF into a universal thermal transport indicator has become essential.

In this work, we introduce a universal macroscopic descriptor ($\sigma$) that enables effective screening of ULG-type $\kappa_L$ in crystalline materials. This descriptor is expressed in a simple form involving only the sound velocity ($v$) and the atomic number of the primitive cell ($n$). Using a large and diverse set of experimental and theoretical data, we established this universal descriptor by identifying a highly correlated Pearson-type relationship and systematically simplifying the Slack model based on fundamental physical quantities. We further integrated this universal descriptor with high-throughput workflows and universal machine-learning interatomic potentials to screen for materials with ultralow $\kappa_L$ across the Materials Project database.[33] We identified two metal halides in the $CsA_2I_3$ (A = Ag, Cu) system as promising ULG-type candidates. We subsequently synthesized $CsAg_2I_3$ and $CsCu_2I_3$ and experimentally confirmed ULG-type $\kappa_L$ values of 0.15–0.16 W/m·K between 170 and 400 K and 0.18–0.20 W/m·K between 300 and 523 K, respectively. Moreover, using a combination of state-of-the-art techniques, including electron diffraction, inelastic neutron scattering, and a unified theory of thermal transport, we comprehensively investigated the relationships among crystal structure, lattice dynamics, and microscopic thermal transport in the representative compound $CsAg_2I_3$. Our work introduces an effective and accessible tool, readily usable by both experimentalists and theoreticians, for screening crystalline materials with targeted ULG-type $\kappa_l$. By establishing a robust correlation between ULG-type thermal transport and laboratory-measurable physical quantities, this study significantly accelerates the discovery of materials for a broad range of thermal-related applications.

## Results and discussion

### Potential indicator for screening ULG-type $\kappa_L$ materials

Using the oversimplified elementary kinetic transport theory, $\kappa_L$ can be computed as $\kappa_L = \frac{1}{3}C_V v_g l$, where $C_v$ is the isochoric heat capacity, $v_g$ is the phonon group velocity, and $l$ is the phonon mean free path. Therefore, a high $v_g$ inherently results in a high lattice thermal conductivity.[34] For instance, Li et al.[34] demonstrated a well-linearized relationship between the minimum thermal conductivity and the $v_g$ for various crystalline materials. Additionally, for crystalline materials with ultralow $\kappa_L$, Agne et al.[27] derived a minimum diffuson-mediated thermal transport model: $\kappa_{L,min} = 0.76(n_d)^{2/3} k_B * \frac{1}{3}(2v_t + v_l)$, where $n_d$ is the atomic number density, $k_B$ is the Boltzmann constant, and $v_t$ and $v_l$ are the transverse and longitudinal sound velocities, respectively. Recently, Wang et al.[26] also derived an interpretable formula for the $\kappa_L$ of crystals: $\kappa_L = \frac{GvV^{1/3}}{nT^\delta} * e^{-r}$, where $G$, $V$, $r$, and are shear modulus, volume of the primitive cell, Grüneisen parameters, and scattering parameter (between 1 and 2). This formula achieves high precision and provides fast predictions. Within the framework of Slack model and some predictive models of minimum $\kappa_L$,[23, 25, 26, 27] various empirical macroscopic quantities are considered as potential correlated parameters influencing ULG-type $\kappa_L$, such as Debye temperature ($\Theta_D$), $n$, volume per atom ($V_a$), density ($\rho$), average mass of atoms ($M_a$), and isobaric heat capacity ($C_p$). To evaluate the correlation between these microscopic quantities, we conducted a Pearson correlation matrix analysis.

To propose a model for assessing ULG-type $\kappa_L$ in crystalline materials, we analyze the relationship between temperature-dependent $\kappa_L$ and $n$. As illustrated in Fig. 1a, the $\kappa_L$ of crystalline materials typically exhibits a universal asymptotic decay, expressed as $\kappa_L(T) \sim T^{-m}$.[35] Specifically, the dominant anharmonic three-phonon scattering mechanisms leads to a trend of $\kappa_L(T) \propto T^{-1}$ ($m=1$),[36] as observed in strongly-bonded materials such as AlN,[37] Si,[38] and GaSb.[39] Strong anharmonic phonon scattering events result in milder decays ($m<1$), as seen in $\kappa_L(T)$ behavior of PbTe ($T^{-0.92}$),[40] InTe ($T^{-0.55}$),[41,42] and SrCuSb ($T^{-0.51}$) compounds.[43] In contrast, in highly disordered or glassy materials, $\kappa_L(T)$ exhibits a temperature independence or even positive temperature dependence, as seen in $Cs_3Bi_2I_9$,[44] $Cu_{12}Sb_4S_{13}$,[45] $Ag_8GeSe_6$,[46]

KCu$_5$Se$_3$,[47] and Cu$_7$PS$_6$.[13] Macroscopically, compounds with a higher number of $n$ generally show a greater tendency to display weak or positive temperature dependence in $\kappa_L$, which can be considered as a critical factor in capturing the glass-like behavior of thermal conductivity.

To qualitatively access the magnitude of ULG-type $\kappa_L$ in crystalline materials, we utilized experimental $\kappa_{avg}$, obtained as the average of the $\kappa_L$ over measured temperatures (Table S1a). We considered the $\kappa_L$ values above Debye temperature for specific crystalline materials, as this temperature range is dominated by phonon-phonon Umklapp scattering. Our Pearson correlation matrix analysis (Fig. 1b) shows that $v$ has the strongest positive correlation value with $\kappa_{avg}$ (0.44), suggesting that ultralow $v$ can serve as a practical and direct parameter for identifying ULG-type $\kappa_L$. Collectively, $\Theta_D$, $v$, anharmonicity, crystal symmetry (designated by the space group number), $n$, $M_a$, $V_a$, $C_p$, number of atoms in the chemical formula, and $\rho$ account for about 98% of the correlation with $\kappa_{avg}$. In contrast, parameters like formation energy and band gap present negligible correlation (2%).

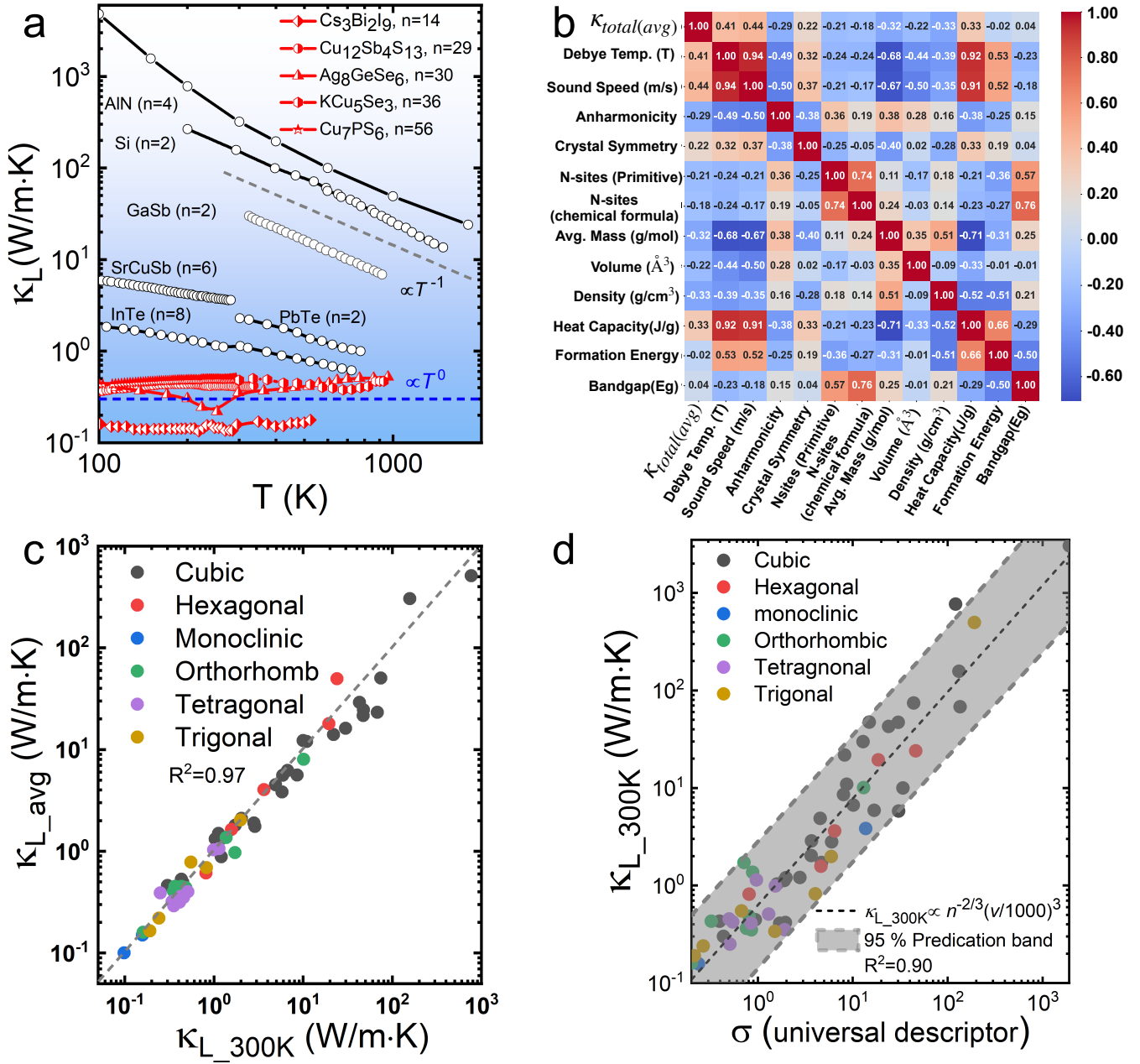

**Fig. 1 Construction of a universal descriptor for screening ULG-type $\kappa_L$. a,** Experimental $\kappa_L$ for various compounds, including Si, AlN, GaSb, PbTe, InTe, SrCuSb, $Cs_3Bi_2I_9$, $Cu_{12}Sb_4S_{13}$, $Ag_8GeSe_6$, $KCu_5Se_3$, $Cu_7PS_6$. **b,** The Pearson correlation matrix calculations of over 40 crystalline materials. **c,** The plot of the experimental $\kappa_{avg}$ vs. $\kappa_{L\_300K}$ of over 60 crystalline samples. **d,** The plot of the experimental $\kappa_{L\_300K}$ vs. $\sigma$ of over 60 crystalline samples. Detailed information on the experimental and calculated data shown in Figs. 1b-d are listed in Table S1a-b.

To simplify the prediction of ULG-type $\kappa_L$ using experimental macroscopic quantities, we begin with a classical Slack model of $\kappa_L$,[48, 49] under dominant Umklapp scattering:

$$\kappa_L \approx A \frac{M_a \Theta_D{}^3 V_a{}^{1/3}}{\gamma^2 n^{2/3} T} \quad (1)$$

where $A$ is a constant. Given that $\gamma$ is not constant but typically varies between 1 and 2 in most crystalline materials, we employ a proportional relationship:

$$\kappa_L \propto \frac{M_a \Theta_D{}^3 V_a{}^{1/3}}{n^{2/3} T} \quad (2)$$

Considering that optical phonons can contribute significantly to thermal conductivity[50, 51] and to develop a universal lattice thermal-conductivity descriptor applicable to both diffuson- and phonon-dominated regimes, we adopt the traditional definition of the Debye temperature, rather than the acoustic-mode Debye temperature. The traditional Debye temperature $\Theta_D$ can be expressed by the relation[52] $\Theta_D = \frac{\hbar v}{k_B}(\frac{3}{4\pi}\frac{n}{V})^{1/3} = \frac{\hbar v}{k_B}(\frac{3}{4\pi}\frac{1}{V_a})^{1/3}$, which we rewrite in the following form:

$$\kappa_L \propto \frac{3\hbar^3}{4\pi k_B{}^3} \frac{M_a v^3}{V_a{}^{2/3} n^{2/3} T} \propto \frac{M_a v^3}{V_a{}^{2/3} n^{2/3} T} \quad (3)$$

Fig. 1c plots the relationship between $\kappa_{L\_300K}$ and $\kappa_{ave}$ (Table S1b), which reveals a strong linear positive correlation. This indicates that crystalline materials with low $\kappa_{L\_300K}$ also tend to have a low $\kappa_{ave}$. Therefore, $\kappa_{ave}$ can be approximated by the following expression:

$$\kappa_{ave} \propto \kappa_{L\_300K} \propto \frac{M_a}{V_a{}^{2/3}} \frac{v^3}{n^{2/3}} \quad (4)$$

After several trials of the possible indicators in formula (4), including $M_a*(v/1000)^3$, $M_a*n^{-\frac{2}{3}}$, $n^{-\frac{2}{3}}(v/1000)^3$, $M_a*(v/1000)^3*n^{-\frac{2}{3}}$, and $M_a*(v/1000)^3*(n*V_a)^{-\frac{2}{3}}$, we identified an indicator $\sigma$, expressed as $n^{-\frac{2}{3}}(v/1000)^3$, that exhibits a linear relation with $\kappa_{avg}$, achieving a goodness of fit of 0.9. Other fitting results between $\kappa_{L\_300K}$ and indicators ($M_a*(v/1000)^3$, $M_a*n^{-\frac{2}{3}}$, $n^{-\frac{2}{3}}(v/1000)^3$, and $M_a*(v/1000)^3*n^{-\frac{2}{3}}$) can be found in Fig. S1. As shown in Fig. 1d, we collected experimental $\kappa_{L\_300K}$, including both ultrahigh and ultralow values, as well as $v$ and $n$ values from over 60 crystalline materials with various crystal structures (Table S1b). $\kappa_{L\_300K}$ typically follows a linear relationship with the established $\sigma$ on a log scale. Our modeling indicates that almost all data points fall within the grey region, representing the 95% of confidence and predication band. This behavior strongly supports the correctness of the proposed indicator $\sigma$, suggesting its effectiveness in identifying ULG-type $\kappa_L$ crystalline materials without the need for time-consuming calculations of $\kappa_L$. However, experimental determination and validation of $v$ still require material synthesis and characterization, which can be time- and energy-intensive for cross-checking.

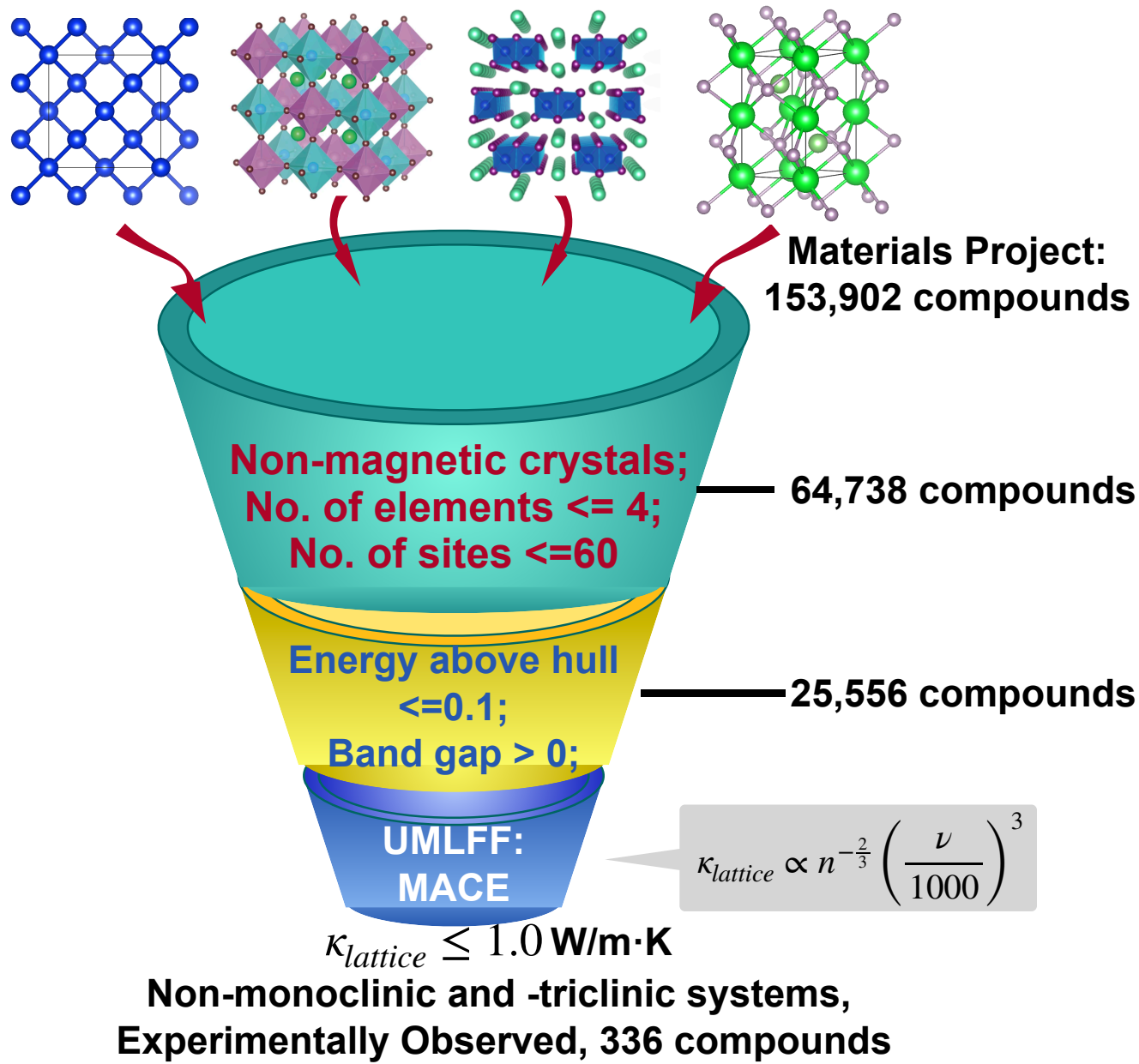

**Fig. 2 Schematic diagram of high-throughput screening process.** High-throughput thermal conductivity calculations using MACE[28] and our universal indicator. The crystal structures were visualized and exported using the VESTA software package.[92]

To overcome this limitation, we conducted high-throughput calculations using a universal machine learning force field method (MACE[28]) to enhance the efficiency and accuracy of $v$ estimations. As illustrated in Fig. 2, we utilized the Materials Project database[33] containing 153,902 materials and applied various screening criteria. These included selecting only non-magnetic compounds, limiting the number of different element types to less than four, restricting the number of atomic sites to less than 60, requiring a band gap greater than 0 eV, and ensuring the energy above the hull was below 0.1 eV/atom. Applying these criteria, we filtered the dataset down to 25,556 materials. We then integrated the universal machine learning force field package MACE[28] with our indicator to predict thermal conductivity. After obtaining the predicted values, we further excluded the low-symmetry systems, i.e., monoclinic and triclinic lattices and consider only crystals identified as experimentally observed in the Materials Project.[48] As a result, we identified only 336 materials with a thermal conductivity below 1.0 W/m·K. Given that metal halides consistently exhibit low thermal conductivity,[11, 44, 51, 53, 54] we selected their analogs, specifically crystalline $CsAg_2I_3$ and $CsCu_2I_3$, from 336 compounds for further experimental investigation.

**Structural evolution and thermal transport properties**

In combination of high-throughput calculations with machine learning techniques (see thee method section), we screened metal halides $CsA_2I_3$ (A = Ag, Cu) and synthesized them for characterizations of their structural and physical properties. Prior studies reported that $CsAg_2I_3$ adopts an orthorhombic crystal structure of *Pnma* space group (No 62),[55, 56] while the structural analog $CsCu_2I_3$ crystallizes in the higher-symmetry *Cmcm* space group (No 63), as plotted in Fig. S2. The corresponding Rietveld refinements of powder X-ray diffraction (PXRD) patterns of the $CsAg_2I_3$ and $CsCu_2I_3$ samples at 300 K confirm their high-purity (Figs. S3-S4). The refined crystallographic data for the two compounds are provided in Table S2-S3. The microscopic morphology and chemical distributions of all the samples, examined using SEM and SEM-EDX, are presented in Figs. S5-S6.

To analyze the phase stability of $CsAg_2I_3$, PXRD patterns were recorded from 250 to 320 K at Cu wavelength with 10 K intervals (Fig. S7). Enlargement of the PXRD patterns reveals that, as the temperature increases, certain diffraction peaks gradually disappear, as exemplified by the peaks observed at 2θ=22.46°, 27.20°, and 28.07° at *T*=250 K. In addition, some peaks shift significantly to lower angles, such as those observed at 2θ=28.83°, 30.68°, and 31.71° at *T*=250 K and at 2θ=28.55°, 30.41°, and 31.43°, respectively, at *T*=320 K. Such peak intensity and position modifications suggest the existence of a structural modification of $CsAg_2I_3$ around room temperature. Analyses (Fig. S8, see details in SI) support the assumption that crystal structure of $CsAg_2I_3$ evolves gradually from 250 to 320 K from its distorted variant crystal structure reported in *Pnma* space group (Fig. S9a-b) to that reported for $CsCu_2I_3$,[57] a crystal structure of higher symmetry in *Bbmm/Cmcm* space group (Fig. S9c-d). Besides, a pronounced minor peak was observed at 278 K in the $C_p$ data (Fig. S10), which probably corresponds to a subtle symmetry modification in CsAg$_2$I$_3$, supporting the results of the PXRD data. The second peak over a broad temperature range of 285-310 K originates from the typical transition of addenda for the low-temperature $C_p$ measurement.

The measured $\kappa_L$ of all sintered polycrystalline samples are displayed in Fig. 3a. The CsAg$_2$I$_3$ sample exhibits a temperature-independent $\kappa_L$ of 0.15-0.16 W/m·K from 170 to 400 K, whereas the CsCu$_2$I$_3$ sample shows values of 0.18-0.20 W/m·K from 300 to 523 K. Notably, compared to other metal halides,[11, 53, 58, 59, 60] the CsA$_2$I$_3$ (A = Ag, Cu) samples exhibit the lowest $\kappa_L$ values across a broad temperature range, validating the practical effectiveness of combining a universal descriptor with high-throughput workflows and universal machine-learning interatomic potentials. Our primary experimental objective is to demonstrate the successful identification of crystalline materials with the ULG-type $\kappa_L$; detailed microscopic analysis of the slight differences in $\kappa_L$ between CsAg$_2$I$_3$ and CsCu$_2$I$_3$ falls outside the scope of this study. To correlate thermal transport properties to crystal structures, we conducted structural and chemical bonding analyses. Since these two samples share structural similarity in the high-temperature *Bbmm/Cmcm* phase, we selected a representative compound, CsAg$_2$I$_3$, for further detailed discussions. The *Bbmm* structure of the 320 K-form of CsAg$_2$I$_3$ (Fig. 3b) consists of [Ag$_2$I$_3$]$^-$ double chains of edge-sharing AgI$_4$ tetrahedrons running along *b* (Fig. 3b), interconnected by (100) layers of Cs$^+$ cations. Remarkably the Ag-I distances, ranging from 2.78 to 2.92 Å, are significantly smaller than that expected from the sum of ionic radii ($r_{Ag^+}$= 1 Å, $r_{I^-}$=2.2 Å). The stability of the structure is clearly ensured by the ionic character of the Cs-I bonds between the [Ag$_2$I$_3$]$^-$ double chains. Each Cs$^+$ cation exhibits eight nearest I$^-$ neighbours with Cs-I distances ranging from 3.93 to 4.18 Å, close to the sum of ionic radii ($r_{Cs^+}$=1.7 Å, $r_{I^-}$=2.2 Å), forming almost regular CsI$_8$ bicapped trigonal prisms sharing faces along *b* and edges along *c*.

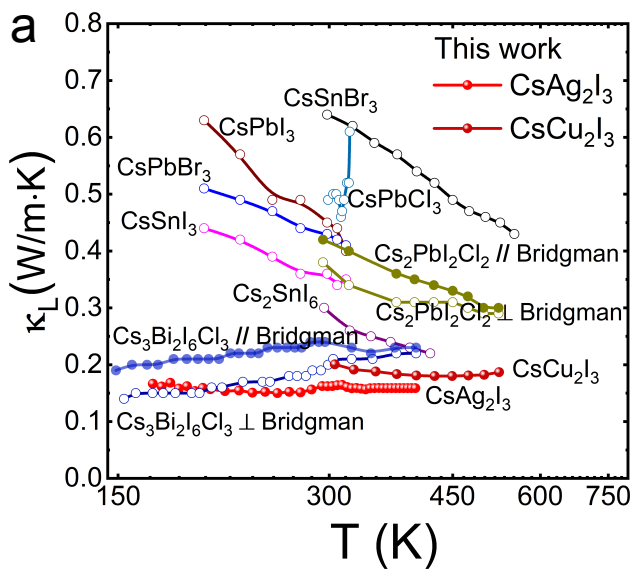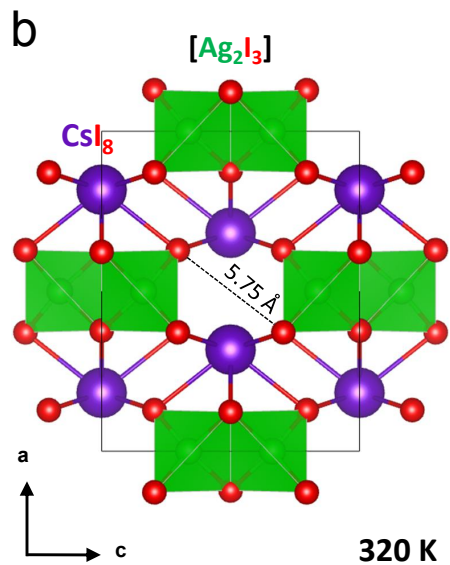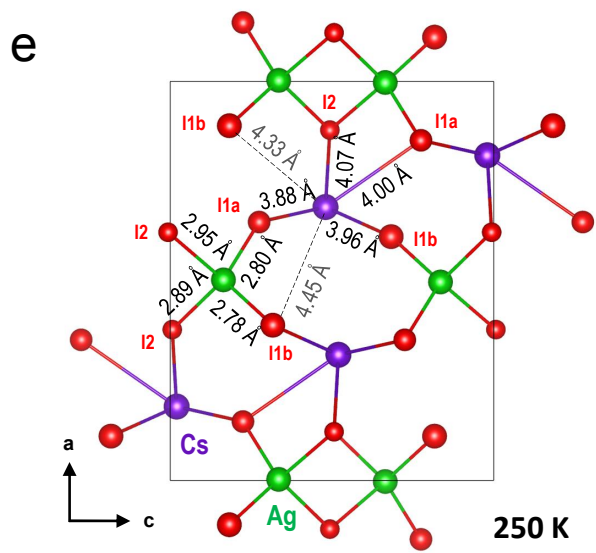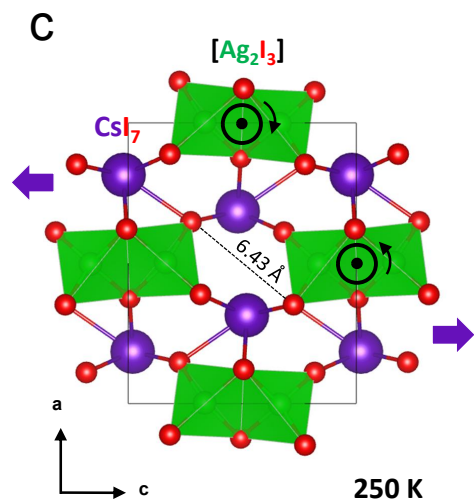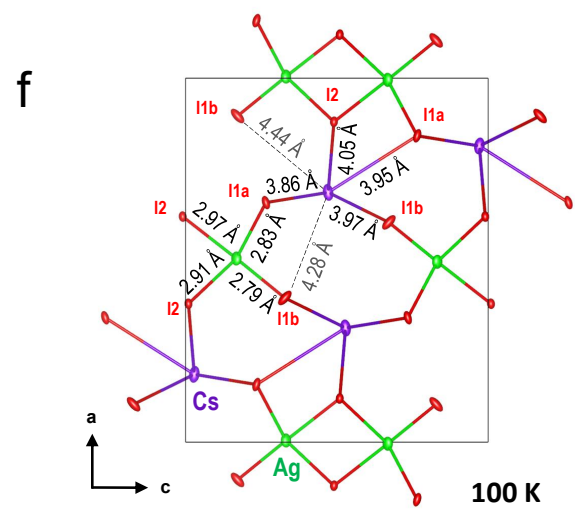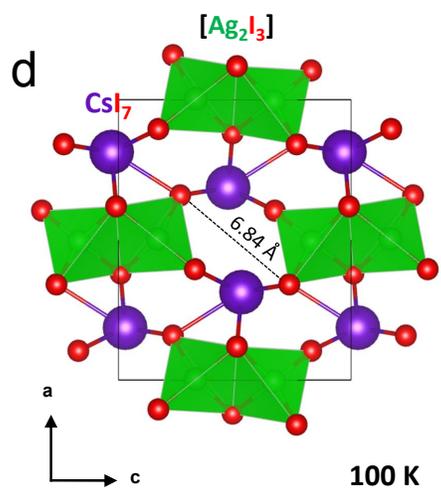

**Fig. 3 Thermal conductivities and structural characterizations of CsA$_2$I$_3$ (A=Ag, Cu) compounds. a,** Comparison $\kappa_L$ of CsAg$_2$I$_3$ and CsCu$_2$I$_3$ with other metal halides compounds.[11, 53, 58, 59, 60] Evolution of the crystal structures of CsAg$_2$I$_3$ with temperature, as determined by **b,** PXRD at 320 K, **c,** PXRD at 250 K, and **d,** 3D ED at 100 K. Crystal structure of CsAg$_2$I$_3$ (SG: *Pnma*) projected along [010], as obtained from **e,** PXRD data recorded at 250 K and **f,** single-crystal 3D ED data recorded at 100 K with main interatomic distances indicated.

The *Pnma* structures of the low temperature form of CsAg$_2$I$_3$ (i.e. 100 K and 250 K, Figs. 3c-d) is very similar to the *Bbmm* form (Fig. 3b). The [Ag$_2$I$_3$]$^-$ double chains of edge-sharing AgI$_4$ tetrahedrons (Fig. 3c) are only slightly distorted with Ag-I distances, ranging from 2.78 to 2.97 Å (Figs. 3e-f). Previously aligned along the *c*-axis in the *Bbmm* form (Fig. 3b), the [Ag$_2$I$_3$]$^-$ double chains are now tilted around the *b*-axis and in antiphase along *a* (Fig. 3c). This tilting is slightly more pronounced at 100 K than at 250 K (Fig. 3c and 3d), in line with the simultaneous increase in the *c* lattice parameter and the decrease in the *a* lattice parameter as the temperature decreases (Fig. S8). It is accompanied by a displacement of Cs$^+$ cations within the (001) plane (Fig. 3c), leading to a significant change in the geometry of the [CsI$_3$] layers (Fig. 3b). Each Cs$^+$ cation now exhibits seven nearest I$^-$ neighbours with Cs-I distances ranging from 3.86 to 4.07 Å (Figs. 3e-f), forming quasi-regular CsI$_7$ monocaped trigonal prisms. Two additional I neighbours corresponding to larger Cs-I distances (4.28 to 4.45 Å) are observed which may not be considered as pure ionic bonds. As a consequence, the structure can be described as (100) [CsI$_3$] layers built up of CsI$_7$ monocaped trigonal prisms sharing faces along *b* and apices along *c* (Fig. 3c).

Such a soft topotactic transition from the high temperature to the low temperature form is explained by the fact that the I(1a) and I(1b) anions in the low temperature *Pnma* form at the border of the [Ag$_2$I$_3$]$^-$ double chains are only linked to two Ag$^+$ cations and consequently can move easily contrary to the I(2) anions located inside the chains which are blocked by 4 Ag$^+$ cations. This is supported by the atomic displacement parameters (ADPs) obtained from both PXRD (Table S4) and 3D ED (Table S8), which show that the I(1a) and I(1b) atoms have higher ADPs than the I(2) atom (Figs. 3e-f). With 3D ED, anisotropic ADPs could be refined, revealing that while I(2) atoms display nearly isotropic ADPs with only slight anisotropy, I(1a) and I(1b) exhibit pronounced anisotropy (Fig. 3f) along a direction associated with the rotation of the [Ag$_2$I$_3$]$^-$ double chains.

This structural analysis emphasizes the key role of iodine vibration in the thermal conductivity of the low temperature form of CsAg$_2$I$_3$ compound. It suggests that I$^-$ governs the appearance of ultralow thermal conductivity in connection with the one dimensional character of the structure. The 1D character of the double [Ag$_2$I$_3$]$^-$ anionic chains facilitates the phonon propagation along the *b* axis, so that higher $\kappa_L^P$ can be expected along this direction. These anisotropic transport properties are moreover reinforced by the fact that the central "Ag-I(2)-Ag" chains form very strong bonds as the I(2) atom is blocked between four $Ag^+$ cations. In contrast, the I(1a) and I(1b) atoms located at the junction between the $CsI_7$ monocapped prisms and the $[Ag_2I_3]^-$ chains are more free due to the fact that they are linked to only two $Ag^+$ cations. Consequently, larger ADPs likely contribute to pronounced anharmonicity at these atomic positions. Such a feature is favourable to the suppression of phonon propagation leading to lower $\kappa_L^P$ values in the directions transversal to the *b*-axis. In the high temperature *Bbmm* form the distribution of the $I^-$ anions is more symmetric, leading to more regular $AgI_4$ tetrahedrons and a more symmetric distribution of the I(1) anionic species (instead of I(1a) and I(1b)). These features are expected to facilitate the phonon transport, increasing $\kappa_L^P$ along the *a*- and *c*-axis. As shown in Table S9a, the experimental average sound velocity ($v_{a,exp}$) of 1188 m/s is considerably lower than that of most thermoelectric materials.[28] The measured $v_{a,exp}$ aligns well with the calculated value of 1125 m/s, as listed in Table S9b, indicating the weak bonding and soft crystal atomic lattices of $CsAg_2I_3$ compound, which might contribute to the low $v_{a,exp}$, reduced elastic properties, significant anharmonicity, and ultimately ultralow $\kappa_L$.

**Lattice dynamics and strong anharmonicity**

For better understand the ultralow $\kappa_L$ of $CsAg_2I_3$, we conducted temperature-dependent inelastic neutron scattering (INS) to study the lattice dynamics properties. Details on the INS technique and data processing are provided in the Supplementary Information (SI). Measurements were performed with two incident energies, $E_i$=12.5 and 30.0 meV, across a wide temperature range of 1.5-300 K, enabling us to analyze the complete generalized density of states (GDOS) as well as the temperature evolution of low-energy phonons. As illustrated in Fig. S11, contour maps of neutron scattering intensity, the dynamic structure factor $S(|Q|, E)$ measured with a higher incident energy $E_i$=30.0 meV, reveals prominent low-energy phonon modes below 5 meV at 1.5 K. Furthermore, we processed the calculated phonon density of states (PDOS) at 0 K by convoluting it with the experimental resolution and applying elemental neutron weighted cross-sections. Note that the calculated neutron weighted PDOS is in close agreement with the experimental GDOS obtained with $E_i$=30.0 meV at 1.5 K (see Fig. 4d).

Fig. 4a depicts $S(|Q|, E)$ measured with $E_i$=12.5 meV at 1.5 K. The measured signal discloses two low-energy phonon bands of high intensity located at about 2 and 3.3 meV and two less intense modes centered around 5 and 7 meV. They are visible as clearly separated peaks in the corresponding GDOS presented in Fig. 4d. The computed element-projected PDOS shown with the experimental data suggests that these low-energy flat bands are dominated by Ag- and I-weighed phonon modes. Interestingly, as indicated by the computed phonon dispersion (blue lines denote acoustic phonon branches; white lines denote optical phonon branches) in Fig. 4a, a strongly dispersive feature can be observed around $|\mathbf{Q}|$=2.5-3.0 Å$^{-1}$. It is characteristic of acoustic phonons. These acoustic phonons exhibit thus an exceptionally low cut-off frequency by crossing the low-energy optical phonons at 1~2 meV. We expect this phonon dispersion to be a favorable configuration for efficient phonon-phonon scattering.

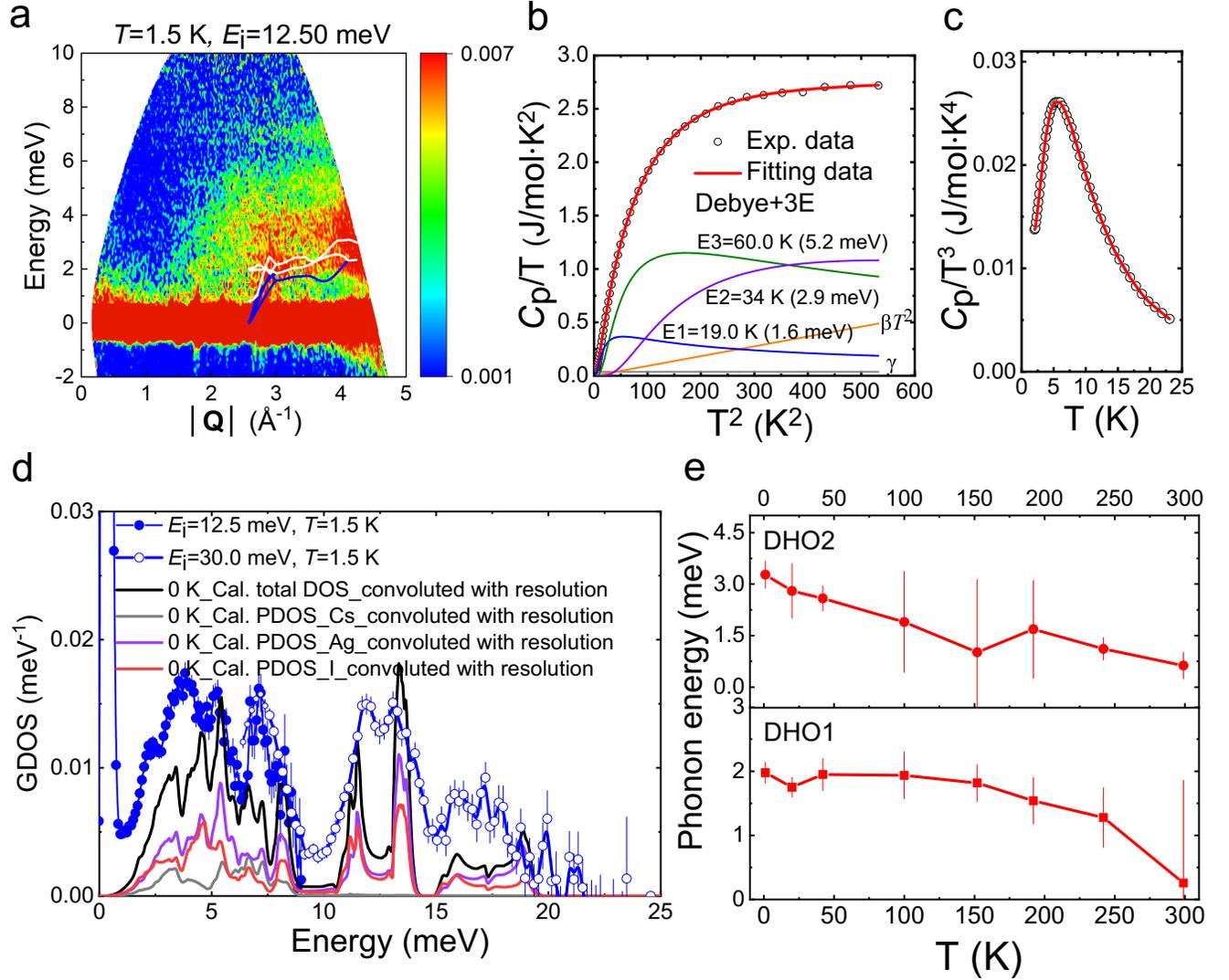

**Fig. 4 Temperature evolution of phonons in CsAg$_2$I$_3$. a,** Experimental $S(|\mathbf{Q}|, E)$ with $E_i$=12.50 meV at $T$=1.5 K. The blue and white solid lines denoted as the acoustic and optical phonons, respectively. **b,** $C_p/T$ vs. $T^2$ and **c,** $C_p/T^3$ vs. $T$ using the Debye-Einstein model. **d,** Experimental GDOS determined with $E_i$=12.5 meV (marked as the blue dotted solid line) and $E_i$=30.0 meV (marked as the blue hollow solid line) at $T$=1.5 K. The calculated PDOS convoluted with a 0.6 meV resolution resulted in a solid black line. The calculated neutron-weighted partial phonon density of states for Cs (grey solid lines), Ag (purple solid lines), and I (pink solid lines) atoms from 1.5-300 K. The experimental data for **e,** temperature-dependent

phonon modes of DHO1 and DHO2. The lines here are guides to the eye. Error bars represent standard deviations.

Additionally, we conducted a fitting analysis of the low-temperature $C_p$ using the Debye-Einstein model. Detailed discussions of the fitting process and the Debye-Einstein model can be found in the referenced study.[13] The $C_p/T$ versus $T^2$ plot in Fig. 4b illustrates a well-fitted profile, indicating three localized low-energy Einstein modes, with values E1=19.0 K (1.6 meV), E2=34.0 K (2.9 meV), and E3=60.0 K (5.2 meV). Importantly, these Einstein modes derived from $C_p$ are also consistent with the INS findings obtained using high incident energy $E_i$=30.0 meV and high-resolution $E_i$=12.5 meV, further supporting the presence of low-energy optical modes in $CsAg_2I_3$. Fig. 4c displays the excellent fitting profile of $C_p/T^3$ vs. T plot with Debye and three Einstein modes from 2 to 23 K. Notably, a distinct Boson peak is observed in the $C_p/T^3$ vs. T plot, revealing the presence of low-energy excitations.[13]

To investigate the temperature dependence of low-energy phonons, we analyzed the constant-momentum spectra at $|Q|$=(3.2±0.1) Å$^{-1}$ (as displayed in Fig. 4a), where the low-energy phonon signal is prominent. These constant $|Q|$ spectra (Fig. S12) were fitted using a combination of functions: four damped harmonic oscillator (DHO) functions,[34] a delta function for the elastic line (convoluted with a Gaussian-shaped resolution of 0.6 meV full-width at half-maximum, FWHM), and a background term. The four DHO components, labeled DHO1 to DHO4 in ascending energy, were used to characterize the phonon modes. The analysis revealed that the characteristic energies of DHO1 and DHO2 undergo continuous softening of the low-energy phonon peaks (Fig. 4e). Specifically, DHO1 and DHO2, with initial energies of 2 and 3.3 meV at 1.5 K, respectively, develop into overdamped phonons as temperature approaches 300 K. These low-energy modes can be viewed as soft phonons, which are closely associated with the displacive phase transition, such as charge-density wave,[35] ferroelectric,[36] and superionic transitions.[37] The appearance of soft phonons in $CsAg_2I_3$ is likely linked to the structural transition from *Pnma* to *Cmcm* around 300 K. Besides, the characteristic energies of the high-energy optical phonons, referred as DHO3 and DHO4, exhibit a phonon softening behavior but without fully softening (Fig. S13) as observed in DHO1 and DHO2. Therefore, the presence of soft phonons and phonon softening in $CsAg_2I_3$ underscores its intrinsically strong anharmonic feature, contributing to its exceptionally low $\kappa_L$.

**Microscopic mechanisms of ultralow glass-like $\kappa_L$**

To better understand thermal transport in crystalline $CsAg_2I_3$ with space groups *Pnma* and *Cmcm*, we calculated $\kappa_L$ using the unified theory of thermal transport,[14] incorporating both population $\kappa_L^P$ and diffuson contributions $\kappa_L^C$, as shown in Fig. 5a. It is worth noting that the calculated $\kappa_L$ of the CsCu$_2$I$_3$ compound was reported[61] while this paper was in preparation, therefore, we focus here only on the theoretical study of the CsAg$_2$I$_3$ compound. When both 3-phonon (3ph) and 4-phonon (4ph) scattering processes are considered, the $\kappa_L^C$ significantly dominates the total $\kappa_L$ along all crystallographic axes. We predict a room-temperature $\kappa_L^C$ of 0.09 W/m·K along the x-axis (0.11 and 0.30 W/m·K along the y- and z-axes, respectively), while the corresponding $\kappa_L^P$ values are 0.02 W/m·K along the x-axis (0.03 and 0.02 W/m·K along the y- and z-axes, respectively) (see Fig. 5a).

Interestingly, even when considering both phonon and diffuson channels, we observe an exceptionally low total $\kappa_L$ of 0.11 and 0.13 W/m·K along the x-axis and y-axis, respectively, which is comparable to the experimentally recorded ultralow $\kappa_L$ of 0.1 W/m·K in crystalline Bi$_4$O$_4$SeCl$_2$.[12] To validate our predicted $\kappa_L$, we averaged the computed total $\kappa_L$ across all crystallographic orientations and compared it with our experimentally measured $\kappa_L$ for the polycrystalline CsAg$_2$I$_3$ sample. From Fig. 5a, we observe that the comparison of $\kappa_L$ between experiment and theory shows good agreement, although the theoretical values are slightly higher than the experimental measurements.

We next examine the glass-like behavior observed in both the experimentally measured and theoretically predicted $\kappa_L$, as illustrated in Fig. 5a. Notably, both the experimentally measured and theoretically predicted $\kappa_L$ exhibit temperature-independent behavior over the range of 170–400 K. This temperature-independent behavior of $\kappa_L$ has also been observed in cubic tetrahedrite Cu$_{12}$Sb$_4$S$_{13}$,[15] although crystalline CsAg$_2$I$_3$ exhibits even lower thermal conductivity. Structural complexity can enhance thermal transport in materials with ultralow thermal conductivity[14] and may cause $\kappa_L$ to exhibit positive temperature dependence[13] due to the dominant role of the diffuson channel.[14] The structural complexity of CsAg$_2$I$_3$, with 12(24) atoms per primitive cell (moderate complexity), may represent a critical threshold to suppress lattice $\kappa_L$ to an ultralow limit.

To gain deeper insight into the microscopic mechanisms of thermal transport in crystalline CsAg$_2$I$_3$, we calculated both the phonon lifetimes and axis-dependent group velocities, as illustrated in Fig. 5b and Fig. S14, respectively. In Fig. 5b, most phonons have lifetimes shorter than the Wigner limit in time,[35] highlighting the dominant role of diffuson contribution in the thermal transport of CsAg$_2$I$_3$, as shown in Fig. 5a. Notably, another eye-catching characteristic is that most phonon lifetimes approach the Ioffe-Regel limit,[62] particularly for phonons with frequencies between 0.2–2.2 THz (approximated 1-9 meV, see Fig. 5b), indicating the strong lattice anharmonicity and main contributor to ultralow $\kappa_L$ in CsAg$_2$I$_3$.

Combined with the structural complexity in crystalline $CsAg_2I_3$, the observation of strong anharmonicity further underscores the minor contribution of the particle-like phonon channel to the total $\kappa_L$. Additionally, from Fig. S14, we observe low $v_g$ in crystalline $CsAg_2I_3$. Specifically, the highest $v_g$ are 2200 m/s along the x-axis (1800 and 1200 m/s along the y- and z-axes, respectively). These low $v_g$ further suppress phonon thermal transport in crystalline $CsAg_2I_3$, consistent with observations reported in the perovskites.[32, 51] This unique characteristics of $v_g$ leads to distinct features in the spectral $\kappa_L^P$ along different crystallographic orientations, as shown in Figs. 5c-d and Fig. S15a.

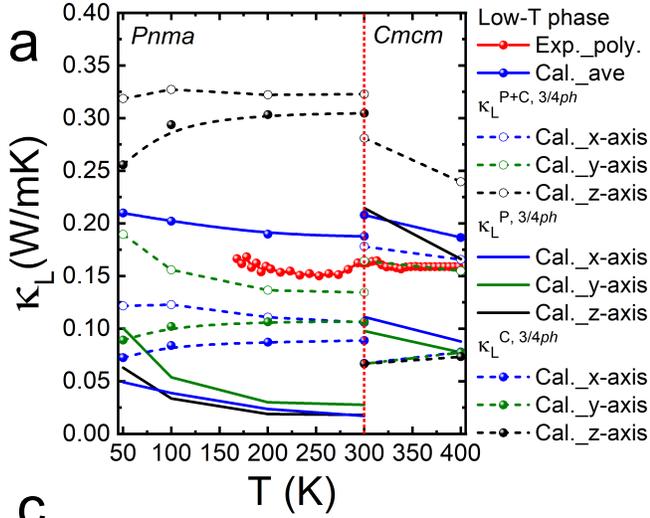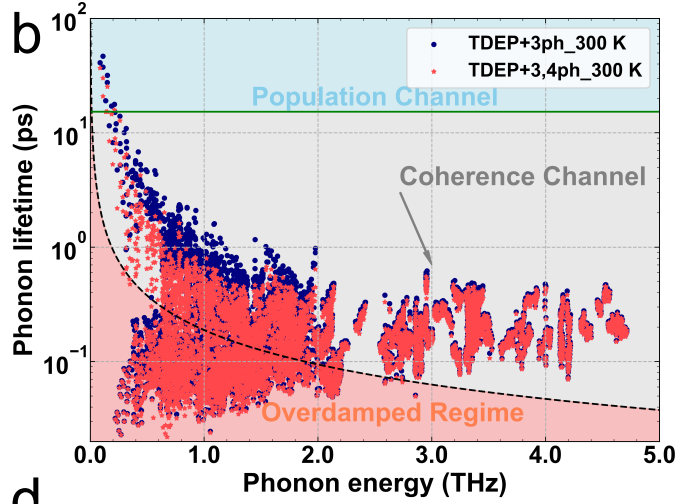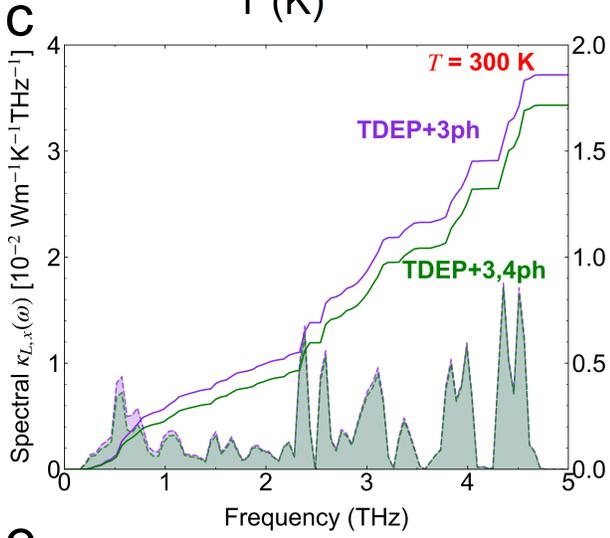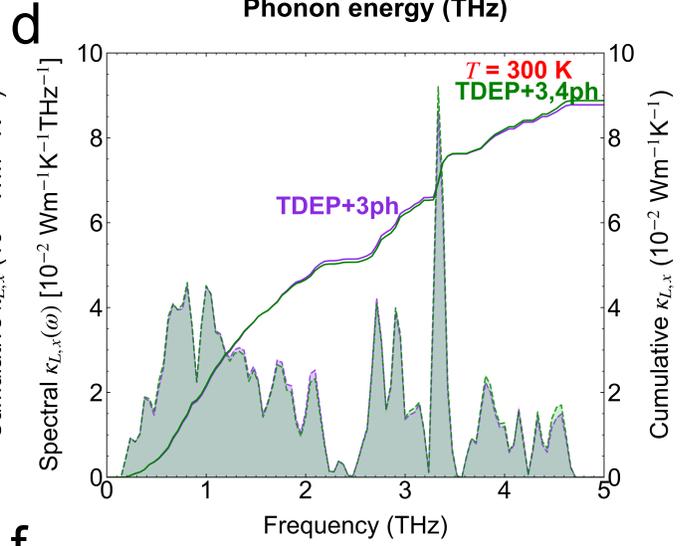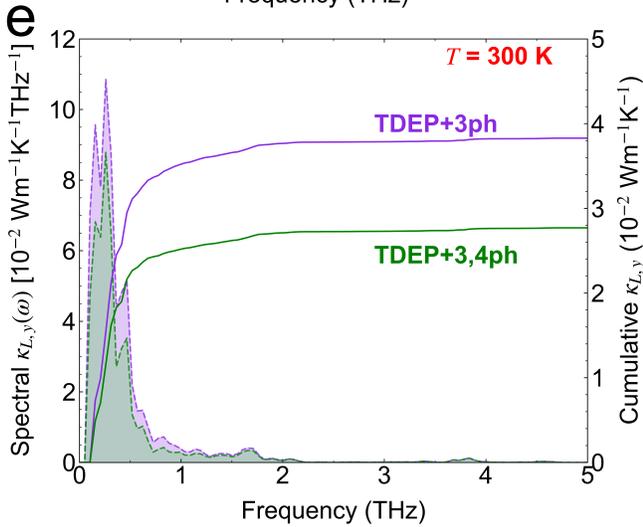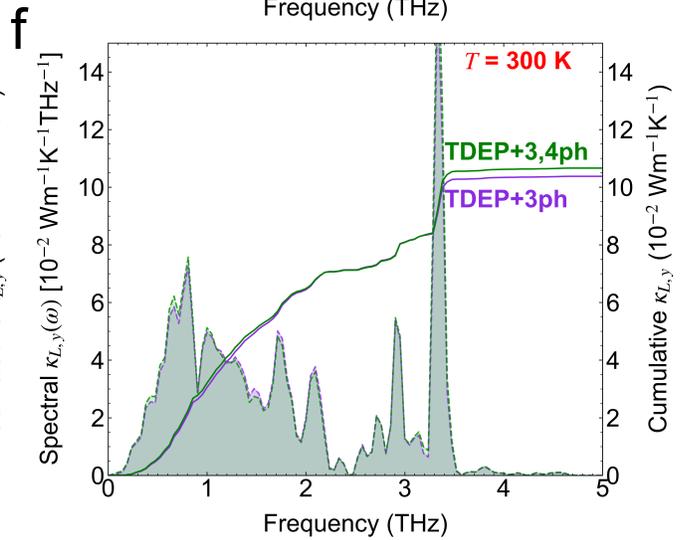

**Fig. 5 Phonon transport properties of CsAg$_2$I$_3$. a,** The calculated axis-dependent $\kappa_\text{L}$, including both population $\kappa_\text{L}^P$ and coherence $\kappa_\text{L}^C$ contributions, for crystalline CsAg$_2$I$_3$ with space group of *Pnma* and *Cmcm*, respectively. Since our experimental $\kappa_\text{L}$ was measured in polycrystalline CsAg$_2$I$_3$, the averaged total thermal conductivity $\kappa_{\text{L,ave}}^{P+C}$ was computed for a fair comparison and shows good agreement with the experimental data. **b,** The calculated phonon lifetime as a function of phonon frequency at *T*=300 K, accounting for 3-phonon processes and both 3- and 4-phonon processes, respectively. The solid green line represents the Wigner limit in time,[14] while the black dashed line indicates the Ioffe-Regel limit in time.[35] **c-d,** The calculated spectral and cumulative population thermal conductivity $\kappa_\text{L}^P(\Omega)/\kappa_\text{L}^P$ along the x- and y-axes at *T*=300 K, considering 3ph and both 3- and 4ph scattering processes, respectively. **e-f,** The same as c-d, but for spectral and cumulative diffuson thermal conductivity $\kappa_\text{L}^C(\Omega)/\kappa_\text{L}^C$.

For instance, the spectral $\kappa_\text{L}^P$ along the x-axis is predominantly contributed by phonons with frequencies ranging from 0 to 4.6 THz (approximated 0-19 meV), whereas the $\kappa_\text{L}^P$ along y- and z-axes is mainly contributed by phonons with frequencies below 1 THz (approximated 4 meV). In contrast, the spectral $\kappa_\text{L}^C$ along all crystallographic orientations is contributed by all phonons, unaffected by the material's structure, due to its diffuson characteristics, as shown in Figs. 5e-f and Fig. S15b. By carefully comparing the atomic participation ratio (APR) (see Fig. S16), phonon lifetimes (see Fig. 5b), and spectral $\kappa_\text{L}^P$ (see Fig. 5c-d and S15a), we find that the strong anharmonicity primarily originates from the I atoms. Notably, the atom-resolved PDOS (see Fig. S17) confirms that the low-energy phonon modes are primarily governed by the I$_2$-related atoms, further corroborating the large atomic vibration observed in the structural analysis. Additionally, the flattening modes associated with Cs atoms contribute significantly to phonon scattering rates. A similar phenomenon has also been observed in perovskites[51] and tetrahedrites.[15] Overall, the combination of weak bonding (low $v_\text{g}$) and strong anharmonicity results in the record-low thermal conductivity observed in crystalline CsAg$_2$I$_3$. Finally, from Fig. 5a, we observe a distinct difference in the phonon and coherence contributions to the total thermal conductivity between the *Pnma* and *Cmcm* phases. In the *Cmcm* phase, phonons contribute significantly to the total thermal conductivity, whereas in the *Pnma* phase, coherence effects dominate. This disparity can be attributed to structural complexity: the *Pnma* phase contains 24 atoms per unit cell, while the *Cmcm* phase contains only 12. The increased structural complexity in *Pnma* suppresses particle-like phonon transport and enhances diffuson contributions to thermal transport.[14, 35]

In summary, informed by the classical Slack model and a unified perspective on minimum $\kappa_\text{L}$, we derived a simple yet powerful universal descriptor, $n^{-\frac{2}{3}}(v/1000)^3$, for rapidly identifying crystalline materials exhibiting ULG-type $\kappa_\text{L}$. To avoid the time- and resource-intensive requirement of experimentally or theoretically determining $v$, we built a global database of $v$ for inorganic crystalline materials using high-throughput calculations and machine-learning methods, thereby accelerating the discovery of materials with targeted thermal transport properties. Guided by the universal descriptor and established global $v$ database, we identified the CsA$_2$I$_3$ (A = Ag, Cu) halide family as a promising materials system, including the compounds CsAg$_2$I$_3$ and CsCu$_2$I$_3$. We synthesized polycrystals and characterized their thermal conductivities, which displayed record ULG-type $\kappa_L$ of 0.15-0.20 W/m·K in the temperature of 170-523 K. To investigate the microscopic relationships between crystal structure, vibrational properties, and phonon transport, we further conducted in-depth PXRD, electron diffraction, INS, and unified theory to a representative compound CsAg$_2$I$_3$.

The temperature-dependent PXRD confirmed the orthorhombic crystal structure of CsAg$_2$I$_3$ characterized by the *Pnma* space group below 300 K and revealed a structural transition to *Cmcm* space group above 300 K. This compound, with a moderate complexity of $n = 24$ for *Pmna* and $n = 12$ for *Cmcm*, can be viewed as consisting of a building unit [Ag$_2$I$_3$]$^-$ anionic chains along the *b*-axis (1D character), along with Cs$^+$ playing the role of counter-cations. The structural analysis emphasizes the key role of iodine vibration and weak Ag-I bonds in the thermal conductivity of the low temperature form, resulting in extremely low $v$ and large anharmonicity. Moreover, our temperature-dependent INS experiments revealed the presence of soft optical phonons located at ~2 meV as well as a phonon softening feature at elevated temperatures, further supporting strong anharmonic feature. Our unified theory of thermal transport also highlights the dominant diffuson thermal transport properties, and the coexistence of extremely low $\kappa_L^C$ and $\kappa_L^p$ in CsAg$_2$I$_3$. These features enabled by moderate complexity and strong anharmonicity, are responsible for ULG-type $\kappa_L$ of this material. Thus, the discovery of record-low ULG-type $\kappa_L$ in the CsA$_2$I$_3$ (A = Ag, Cu) halide system provides compelling evidence for the utility of our universal descriptor, high-throughput methodology, and universal machine-learning potentials in identifying crystalline materials exhibiting ULG-type $\kappa_L$. The applicability and versatility of our descriptor are anticipated to bridge the gap in evaluating ultralow $\kappa_L$ in crystalline inorganic materials across a broad temperature range. By offering a low-cost and easily deployable approach for both experimentalists and theorists, this work enables accelerated discovery of thermally functional materials for diverse applications.

## Methods

**Synthesis**

The metal halides compounds CsAg$_2$I$_3$ and CsCu$_2$I$_3$ were prepared using stoichiometric amounts of precursors CsI (powder, 99.9%), AgI (powder, 99.9%), and CuI (powder, 99.9%). The mixtures of precursors were ground into fine powders in an Ar-filled glove box. The resulting powders were then cold-pressed into pellets and placed in carbon coated silica tubes. These tubes were subsequently evacuated and sealed under a vacuum of ~10$^{-3}$ Pa. The sealed tubes were heated to 623 K over a period of 7 hours and held at this temperature for an additional 24 hours, followed by cooling down within 7 hours. The obtained ingots of CsAg$_2$I$_3$ and CsCu$_2$I$_3$ were ground into fine powders and loaded into a 10 mm graphite die, followed by densification through spark plasma sintering (SPS) at 473 K and 523 K for 5 minutes under an applied pressure of 64 MPa. The resultant SPS-ed samples were confirmed to be a highly densified pellet with a 98% of the theoretical density.

**X-ray diffraction**

Temperature-dependent high-resolution PXRD of the synthesized SPS-ed CsAg$_2$I$_3$ powder were conducted from 250 to 320 K utilizing a Smartlab 9kW Phenix chamber system equipped with Cu K$_{\alpha 1}$ radiation (λ=1.5406 Å) and Johansson Ge (111) monochromator. High-resolution PXRD data of the synthesized SPS-ed CsCu$_2$I$_3$ powders were recorded at 300 K using a Bruker D8 Advance Vario 1 two-circle diffractometer (θ−2θ, Bragg−Brentano mode) equipped with Cu K$_{\alpha 1}$ radiation (λ=1.5406 Å) and a Ge (111) monochromator (Johansson type). Rietveld refinements were carried out using the FullProf[63] and WinPlotr[64] software packages, incorporating the refinement of parameters such as zero-point shift, unit cell, peak shape, asymmetry, atomic coordinates, and $B_{iso}$ values of each atoms.

**3D electron diffraction and scanning electron microscopy**

3D Electron Diffraction (3D ED)[65] is a technique that uses electron diffraction patterns collected from a single crystal to determine its crystal structure. For this study, precession-assisted 3D ED data were collected using a JEOL F200 transmission electron microscope at 200 kV, equipped with an ASI Cheetah M3 detector and a Nanomegas Digistar precession unit. A CsAg$_2$I$_3$ pellet was ground in ethanol with an agate mortar, and a drop of the resulting suspension was deposited onto a holey carbon membrane on a Cu mesh grid. 3D ED data were recorded at 100 K using a GATAN ELSA tomography holder and the Instamatic program,[66] with a precession angle of 1.25° and a tilt step of approximately 2° between patterns.

Data processing was performed with PETS2,[67] and structure refinements were carried out in Jana2020,[68] accounting for both dynamical diffraction effects and electron beam precession. The refined structural parameters of $CsAg_2I_3$ are summarized in Table S8. Scanning electron micrographs and electron energy dispersive spectroscopy (EDS) analyses of samples were performed using a JEOL JSM-7200F-SEM equipped with an EDX X-Flash Bruker detector.

**Heat capacity and thermal conductivity measurements**

The thermal conductivity ($\kappa$) was calculated employing the formula $\kappa = \rho C_p d$. The wide-range thermal diffusivity ($d$) of the $CsAg_2I_3$ sample was measured using a Netzsch LFA 467 Hyper flash system under a nitrogen atmosphere from 170 to 400 K, while high-temperature $d$ of the $CsCu_2I_3$ samples were conducted utilizing a Netzsch LFA 457 laser flash system under a nitrogen atmosphere from 300 to 523 K. The density ($\rho$) was determined via the Archimedes method. $C_p$ measurement of the $CsAg_2I_3$ sample were conducted in the temperature range of 2 to 350 K using a conventional relaxation method with the dedicated $^4He$ option of the PPMS.

**Sound velocity measurements**

The pulse-echo method was employed to measure the longitudinal and transverse sound velocities of the $CsAg_2I_3$ sample at 300 K. A small quantity of grease was applied to ensure effective contact between the sample and the piezoelectric transducers.

**Inelastic neutron scattering**

Inelastic neutron scattering experiments were conducted on a powder sample weighing ~10 g. The sample was measured at temperatures between 1.5 and 300 K using the thermal neutron time-of-flight spectrometer PANTHER at the Institut Laue Langevin in Grenoble, France. We employed nominal incident energies $E_i$ of 30.0 meV to examine the full phonon spectrum and 12.5 meV to monitor low-energy phonons. We utilized a standard cryostat, with the powder sample contained in an aluminum sample holder. The acquisition periods varied from 1 to 2 hours, depending on $E_i$ and temperature, which influenced the inelastic intensity.

The evolution of the low-energy phonons was recorded during cooling from 300 K to 1.5 using an $E_i$ = 12.5 meV. An acquisition period of 10 minutes led to a temperature difference of 5 K or less between spectra. To ensure accurate data analysis, we performed additional measurements on empty sample holders, the sample environments, and vanadium. These measurements were essential for standardizing our data correction procedure and converting the corrected signal to the dynamic structure factor $S(|\mathbf{Q}|, E)$ and the GDOS $g^{(n)}(E)$. We utilized the Mantid software package for the correction and conversion computing.[69]

For a given polyatomic material, GDOS can be approximated from the atomic partial phonon density of states ($g_i^{(n)}(E)$) of element i as equation (1):[70]

$$g^{(n)}(E) = \sum_i f_i \frac{\sigma_i}{m_i} g_i(E) \exp(-2W_i) \quad (5)$$

Where $i$, $f_i$, $\sigma_i$, $m_i$, and $W_i$ denote the different elements, atomic concentration, neutron total cross-section, mass, and Debye-Waller factor of element $i$, respectively. The values of $\frac{\sigma_i}{m_i}$ for Cs, Ag, and I are 0.03, 0.046, and 0.03 barn/amu.

**Density functional theory calculations**

In this work, all the *ab initio* calculations for crystalline CsAg$_2$I$_3$ were conducted using density functional theory (DFT),[71] as implemented in the Vienna Ab initio Simulation Package (VASP).[72] The projector-augmented wave (PAW)[73] pseudopotentials were utilized to explicitly treat the $5s^25p^66s^1$, $4d^{10}5s^1$ and $5s^25p^5$ electrons as valence states for Cs, Ag, and I atoms, respectively. For the exchange-correlation functional in all DFT calculations, we employed the revised Perdew-Burke-Ernzerhof (PBE) version for solids, i.e., PBEsol functional,[74] within the generalized gradient approximation (GGA) framework.[75] In both structural relaxation and self-consistent DFT calculations, tight convergence criteria were applied, with force convergence set to $10^{-5}$ eV·Å$^{-1}$ and energy convergence to $10^{-8}$ eV. For crystalline CsAg$_2$I$_3$ with a space group of *Pnma*, a kinetic energy cutoff of 600 eV and a Γ-centered 8×10×8 Monkhorst-Pack *k*-mesh were used to sample the Brillouin zone in the primitive cell, which contains 24 atoms. The resulting fully relaxed lattice constants are $a$ = 13.9250 Å, $b$ = 5.8845 Å and $c$ = 11.3099 Å. Similarly, for crystalline CsAg$_2$I$_3$ with a space group of *Cmcm*, the same kinetic energy cutoff and a Γ-centered 10×10×10 Monkhorst-Pack *k*-mesh were employed to sample the Brillouin zone in the primitive cell containing 12 atoms. The fully optimized lattice constants are ($a$ = $b$ = 8.9727 Å and $c$ = 5.8862 Å).

**Effective harmonic force constants extraction**

To obtain the effective harmonic phonon frequency at finite temperatures, the temperature-dependent effective potential (TDEP) method[76] was used to fit first-principles forces to an effective Hamiltonian ($H$),

$$H = U_0 + \sum_i \frac{P_i^2}{2m_i} + \frac{1}{2}\sum_{ij\alpha\beta} \Phi_{ij}^{\alpha\beta} u_i^\alpha u_j^\beta, \tag{6}$$

where $U_0$ denotes the potential energy, $i$ and $j$ denote the atomic indeces, $p_i$ denotes the momentum associated with atom $i$, $m_i$ denotes atomic mass, $u_i$ denotes atomic displacment. $\Phi_{ij}^{\alpha\beta}$ denotes the effective harmonic interatomic force constants (IFCs), associated with the Cartesian indices $\alpha$ and $\beta$.

To prepare the finite-temperature displacement-force dataset for IFCs extraction, stochastic sampling of the canonical ensemble was employed to generate perturbed supercells,[77, 78] followed by precise self-consistent DFT calculations to determine the forces on all atoms. The Cartesian displacement ($u_i^\alpha$) is normally distributed around the mean thermal dispacement and is expressed as

$$u_i^\alpha = \sum_q e_q^{i\alpha} \langle A_{iq} \rangle \sqrt{-2ln\zeta_1} \sin(2\pi\zeta_2), \tag{7}$$

with the thermal amplitude $\langle A_{iq} \rangle$ given as[77, 78, 79]

$$\langle A_{iq} \rangle = \sqrt{\frac{\hbar(2n_q^0+1)}{2m_i\omega_q}}, \tag{8}$$

where $q$ represents the phonon mode, acting as a composite index that combines the wavevector **q** and phonon branch $s$, $e_q$ represents the eigenvector, $\zeta_1$ and $\zeta_2$ represent stochastically sampled numbers between 0 and 1, $\hbar$, $n_q$ and $\omega_q$ represent the Planck constants, the occupation number following the Bose-Einstein distribution and the phonon frequency, respectively.

In this work, a 1×2×2 supercell containing 96 atoms of crystalline CsAg$_2$I$_3$ with space group of $Pnma$ was used to perform calculations iteratively, starting with 500 thermally perturbed snapshots. The resulting 500 atomic configurations were then used to generate the displacement-force dataset through precise DFT calculations with a Γ-centered 4×4×4 Monkhorst-Pack k-point density grid. Each iteration involves key procedures, including computing phonon normal modes, generating perturbed snapshots, calculating precise DFT forces, and fitting effective IFCs. To ensure the convergence of finite-temperature IFCs, the final iteration at each temperature (50, 100, 200, 300, and 400 K) was performed using 1,500 snapshots.

For the crystalline CsAg$_2$I$_3$ with space group of $Cmcm$, a 2×2×2 supercell containing 96 atoms was used for the iterative calculations of effective IFCs. Similarly, the iteration was initiated with 500 thermally perturbed snapshots generated from zero-K phonon dispersions. Precise forces were obtained through self-consistent DFT calculation using a Γ-centered 4×4×4 Monkhorst-Pack k-point density grid. The final iteration at each temperature (300 and 400 K) was conducted with 1,500 snapshots to ensure the convergence of finite-temperature IFCs. In this study, the temperature-dependent effective potential calculations were performed using the ALAMODE package.[80]

**Effective anharmonic force constants extraction**

Instead of the least-squares approach for extracting the effective harmonic interatomic force constants (IFCs), the Compressive Sensing Lattice Dynamics (CSLD) method[81] was employed to efficiently and accurately extract the anharmonic IFCs. The CSLD method efficiently identifies the physically significant terms from a large set of irreducible anharmonic IFCs using a limited displacement-force dataset.[81] To extract the effective anharmonic IFCs, we sampled a dataset consisting of 300 configurations, with each configuration containing up to 96 displacement-force pairs, from the previous harmonic canonical ensemble at finite temperatures.[76, 82] Subsequently, the effective harmonic IFCs and the displacement-force dataset were used as an input to extract anharmonic IFCs up to the sixth order. The anharmonic IFCs of crystalline $CsAg_2I_3$ with both space groups *Pnma* and *Cmcm* were extracted using the least absolute shrinkage and selection operator (LASSO) technique,[83] with real-space cutoff radii of 7.41 Å, 6.35 Å, 4.23 Å, and 3.17 Å for the cubic, quartic, quintic, and septic IFCs, respectively. In this work, the IFCs fitting process was performed using the ALAMODE package.[80, 81, 84]

**Anharmonic phonon scattering rates**

With the effective interatomic force constants (IFCs) available and using the Fermi's golden rule,[85] the phonon scattering rates for three-phonon (3ph) $\Gamma_q^{3ph}$ and four-phonon (4ph) $\Gamma_q^{4ph}$ processes under the single-mode relaxation time approximation (SMRTA) are expressed as[15, 85]

$$\Gamma_q^{3ph} = \sum_{q'q''} \left\{ \frac{1}{2}(1 + n_{q'}^0 + n_{q''}^0)\mathcal{L}_- + (n_{q'}^0 - n_{q''}^0)\mathcal{L}_+ \right\}, \qquad (9)$$

$$\Gamma_q^{4ph} = \sum_{q'q''q'''} \left\{ \frac{1}{6}\left(\frac{n_{q'}^0 n_{q''}^0 n_{q'''}^0}{n_q^0}\right)\mathcal{L}_{--} + \frac{1}{2}\left(\frac{(1+n_{q'}^0)n_{q''}^0 n_{q'''}^0}{n_q^0}\right)\mathcal{L}_{+-} + \frac{1}{2}\left(\frac{(1+n_{q'}^0)(1+n_{q''}^0)n_{q'''}^0}{n_q^0}\right)\mathcal{L}_{++} \right\}, \qquad (10)$$

with

$$\mathcal{L}_{\pm} = \frac{\pi\hbar}{4N}\left|V^{(3)}(q,\pm q',-q'')\right|^2 \Delta_{\pm} \frac{\delta(\Omega_q \pm \Omega_{q'} - \Omega_{q''})}{\Omega_q \Omega_{q'} \Omega_{q''}}, \tag{11}$$

and

$$\mathcal{L}_{\pm\pm} = \frac{\pi\hbar^2}{8N^2}\left|V^{(4)}(q,\pm q',\pm q'',-q''')\right|^2 \Delta_{\pm\pm} \frac{\delta(\Omega_q \pm \Omega_{q'} \pm \Omega_{q''} - \Omega_{q'''})}{\Omega_q \Omega_{q'} \Omega_{q''} \Omega_{q'''}}, \tag{12}$$

where $\Omega_q$ denotes the effective phonon frequency, $V^{(3)}(q,\pm q',-q'')$ and $V^{(4)}(q,\pm q',\pm q'',-q''')$ denote the reciprocal forms of 3rd- and 4th-order effective IFCs,[86] i.e., the strength of the 3ph and 4ph scattering matrices, respectively. In 3ph and 4ph scattering processes, energy conservation is enforced using the delta function $\delta$, while momentum conservation is imposed via the Kronecker delta $\Delta$.

The extrinsic phonon-isotope scattering rate $\Gamma_q^{iso}$ also plays a critical role in thermal transport and can be expressed as[87]

$$\Gamma_q^{iso} = \frac{\pi \Omega_q^2}{2N} \sum_{i \in u.c.} g(i) \left|e_q^*(i) \cdot e_{q'}(i)\right|^2 \delta(\Omega - \Omega'), \tag{13}$$

Where $g(i)$ denotes the Pearson deviation coefficient. Under the assumption of Matthiessen's rule, the total phonon scattering rate $\Gamma_q$ for a specific mode $q$ is expressed as

$$\Gamma_q = \Gamma_q^{3ph} + \Gamma_q^{4ph} + \Gamma_q^{iso}, \tag{14}$$

**The unified theory of thermal transport**

Given the strong anharmonicity and structural complexity of crystalline CsAg$_2$I$_3$, the unified theory of thermal transport proposed by Michelle *et al.*,[14, 35] which accounts for both $\kappa_L^C$ and $\kappa_L^P$, was employed to calculate the total lattice thermal conductivity $\kappa_L$. Within the SMRTA framework, the unified theory of thermal transport is expressed as[14, 35]

$$\kappa_L^{P/C} = \frac{\hbar^2}{k_B T^2 V N} \sum_q \sum_{s,s'} \frac{\Omega_{qs} + \Omega_{qs'}}{2} \cdot v_{qss'} \otimes v_{qs's} \cdot \frac{\Omega_{qs} n_{qs}(n_{qs}+1) + \Omega_{qs'} n_{qs'}(n_{qs'}+1)}{4(\Omega_{qs} + \Omega_{qs'})^2 + (\Gamma_{qs} + \Gamma_{qs'})^2} \cdot (\Gamma_{qs} + \Gamma_{qs'}), \tag{15}$$

where $k_B$, $T$, $V$, and $N$ are the Boltzmann constant, the cell volume, the absolute temperature and the number of sampled phonon wave vectors, $v$ is the inter- and intra-band group velocity matrix and can be expressed as[88]

$$v_{qss'} = \frac{\langle e_{qs} | \frac{\partial D(q)}{\partial q} | e_{qs'} \rangle}{2\sqrt{\Omega_{qs}\Omega_{qs'}}}. \tag{16}$$

Where $D(q)$ denotes the dynamical matrix. The interband term ($s = s'$) results in diagonal terms of heat flux operators and contributing to $\kappa_L^P$. In contrast, the intraband term ($s \neq s'$) gives rise to off-diagonal terms of heat flux operators and contributing to $\kappa_L^C$. To calculate the total lattice thermla conductivity $\kappa_L$ of crystalline $CsAg_2I_3$ with space group of $Pnma$($Cmcm$), which contains 24(12) atoms in the primitive cell, a $q$ mesh of 8×8×8 (12×12×12) was used for both 3ph and 4ph scattering processes, with a scalebroadening parameter of 0.1(0.2). The parameters chosen for Eq.(15) were thoroughly tested for the convergence and yield well-converged results. In this work, lattice thermal conductivity calculations, including population and coherence contributions, were performed using the ShengBTE[89] and FourPhonon[86] packages, along with our in-house code.[50, 51]

**The high-throughput calculation and universal machine-learning force field: MACE**

MACE is an equivariant message passing neural network potential, using higher order messages combined with Atomic Cluster Expansion (ACE), a method for deriving an efficient basis to represent functions of atomic neighborhoods. Note that the version of MACE used in this work is the MACE-MP-0 (medium) (https://github.com/ACEsuit/mace-foundations), as implemented in the atomate2 package.[90]

**Data availability**

The data supporting the construction of a universal descriptor of this study are all available in the main text and supplementary information. The high-throughput calculation of group velocity was obtained using MACE,[28] and the original data generated in this study is available in the repository: https://github.com/leslie-zheng/High-throughput-group-velocity-calculation. The raw data of the inelastic neutron scattering experiments is available in the repository: 10.5291/ILL-DATA.7-01-592.

**Code availability**

The open-source codes can be found as following: Alamode is available at https://github.com/ttadano/alamode, ShengBTE is available at https://www.shengbte.org, and FOURPHONON is available at https://github.com/FourPhonon/FourPhonon. Atomate2 is available at https://github.com/materialsproject/atomate2. MACE is available at https://github.com/ACEsuit/mace. The in-house codes will be available from the corresponding authors upon reasonable request.

**Acknowledgements**

X.S. acknowledges funding supported by National Youth Talent Project (Overseas) and the Fundamental Research Funds for the Central Universities (D5000250021), and the European Union's Horizon 2020 research and innovation program under the Marie Sklodowska-Curie grant agreement No 101034329 and the WINNING Normandy Program. X.S. and M.M.K acknowledge the European Neutron Source Institut

Laue Langevin (ILL), Grenoble, France for the provision of neutron beamtime of our proposal (No.7-01-592)[91] at beamline Panther. We thank Geoffroy Hautier (Rice University, formerly at Dartmouth College) for thoughtful review of the manuscript and his insightful suggestions. We thank Hrushikesh Sahasrabuddhe (University of California, Berkeley) for his valuable discussions on the machine-learning calculations. We also thank Dr. Mani Jayaraman (Norwestern Polytechinical University) for his assistance with data collection.

**Author contributions**

X.S. conceived the idea, and X.S. and J.Z. designed the project. Materials synthesis and characterizations: X.S.; P.L.; J.H.; P.B.; B.R.; J.W.; J.L.; P.L.; C.C.; E.G.; Theory and modeling: X.S.; J.Z.; Inelastic neutron scattering measurements and analysis: X.S.; M.M.K; DFT and high-throughput calculations: J.Z.; Manuscript draft: X.S; J.Z. with input from the other co-authors; Review and editing: X.S.; J.Z. and other co-authors.

**Competing interests**

The authors declare no competing interests.